%% file: Single_Photon_Transistor_Mediated_by_Inter-State_Rydberg_Interaction.tex
\documentclass[aps,prl,twocolumn,floatfix,amsmath,superscriptaddress,amssymb,bibnotes]{revtex4}

\usepackage{graphicx}
\graphicspath{ {./fig/} }
\usepackage{dcolumn}
\usepackage{bm}
\usepackage{epstopdf}
\usepackage[]{SIunits}
\usepackage{units}
\usepackage{braket}
\usepackage{pbox}

\usepackage[colorlinks, linkcolor = blue, citecolor = blue, filecolor = black, urlcolor = blue]{hyperref}

\begin{document}

\title{Single Photon Transistor Mediated by Inter-State Rydberg Interaction}

\author{H. Gorniaczyk}
\email[]{h.gorniaczyk@physik.uni-stuttgart.de}
\affiliation{5. Physikalisches Institut, Universit\"{a}t Stuttgart, Pfaffenwaldring 57, 70569 Stuttgart, Germany}
\author{C. Tresp}
\affiliation{5. Physikalisches Institut, Universit\"{a}t Stuttgart, Pfaffenwaldring 57, 70569 Stuttgart, Germany}
\author{J. Schmidt}
\affiliation{5. Physikalisches Institut, Universit\"{a}t Stuttgart, Pfaffenwaldring 57, 70569 Stuttgart, Germany}
\author{H. Fedder}
\affiliation{3. Physikalisches Institut, Universit\"{a}t Stuttgart, Pfaffenwaldring 57, 70569 Stuttgart, Germany}
\author{S. Hofferberth}
\email[]{s.hofferberth@physik.uni-stuttgart.de}
\affiliation{5. Physikalisches Institut, Universit\"{a}t Stuttgart, Pfaffenwaldring 57, 70569 Stuttgart, Germany}


\begin{abstract}
We report on the realization of an all-optical transistor by mapping gate and source photons into strongly interacting Rydberg excitations with different principal quantum numbers in an ultracold atomic ensemble. We obtain a record switch contrast of $\unit[40]{\%}$ for a coherent gate input with mean photon number one and demonstrate attenuation of source transmission by over 10 photons with a single gate photon. We use our optical transistor to demonstrate the nondestructive detection of a single Rydberg atom with a fidelity of 0.72(4).
\end{abstract}

\maketitle

In analogy to their electronic counterparts, all-optical switches and transistors are required as basic building blocks for both classical and quantum optical information processing \cite{Dolev2010,Vuckovic2009}. Reaching the fundamental limit of such devices, where a single gate photon modifies the transmission or phase accumulation of multiple source photons, requires strong effective interaction between individual photons. Engineering sufficiently strong optical nonlinearities to facilitate photon-photon interaction is one of the key goals of modern optics and immense progress towards this goal has been made in a variety of systems in recent years. Most prominent so far are cavity QED experiments, which use high finesse resonators to enhance interaction between light and atoms \cite{Kimble2005, Rempe2007, Rempe2008b, Vuletic2011, Rauschenbeutel2013} or artificial atoms \cite{Imamoglu2000,Vuckovic2007,Vuckovic2008,Imamoglu2012}. In particular, a high-contrast, high-gain optical transistor operated by a single photon stored in an atomic ensemble inside a cavity has recently been demonstrated \cite{Vuletic2013}. Cavity-free approaches include single dye molecules \cite{Sandoghdar2009} and atoms coupled to hollow-core \cite{Lukin2009,Gaeta2011} and tapered nano-fibers \cite{Rauschenbeutel2010}.

A novel free-space approach to realize single-photon nonlinearities is to map strong interaction between Rydberg atoms \cite{Saffman2010} onto slowly travelling photons \cite{Adams2010,Vuletic2012} using electromagnetically induced transparency (EIT) \cite{Fleischhauer2005}. This has already been used to demonstrate highly efficient single-photon generation \cite{Kuzmich2012b}, attractive interaction between single photons \cite{Vuletic2013b}, entanglement generation between light and atomic excitations \cite{Kuzmich2013}, and most recently single-photon all-optical switching \cite{Duerr2014}.

However, demonstration of amplification, that is, controlling many photons with a single one, has so far only been achieved in a cavity QED setup \cite{Vuletic2013}. Gain $G\gg 1$ is one of the key properties of the electric transistor that lies at the heart of its countless applications. In this work, we demonstrate an all-optical transistor with gain $G>10$. We employ two-color Rydberg EIT in a free-space ensemble to independently couple gate and source photons to different Rydberg states \cite{Lukin2011}. For a coherent gate pulse containing on average one photon, we observe a switch contrast of $C_\text{coh} = 0.39(4)$, defined as
\begin{equation}\label{eq:Contrast}
C = 1 - \left(\overline{N}_\text{s,out}^\text{with gate} /~ \overline{N}_\text{s,out}^\text{no gate}\right) \text{,}
\end{equation}
where $\overline{N}_\text{s,out}^\text{with gate}$ and $\overline{N}_\text{s,out}^\text{no gate}$ denote the mean numbers of transmitted source photons with and without gate photons. From this we extrapolate that a single-photon Fock state gate pulse causes a switch contrast $C_{sp} = 0.53(2)$. The measured switch contrast is robust against the number of incoming source photons, enabling us to attenuate the source transmission by 10 photons with a coherent gate photon input with mean photon number $\overline{N}_\text{g,in} =0.75(3)$. Finally, we use our transistor to implement single-shot non-destructive detection of single Rydberg atoms in a dense ultracold ensemble \cite{Weidemueller2012,Lesanovsky2011}.
%
%
\begin{figure}[!b]
\begin{center}
\includegraphics{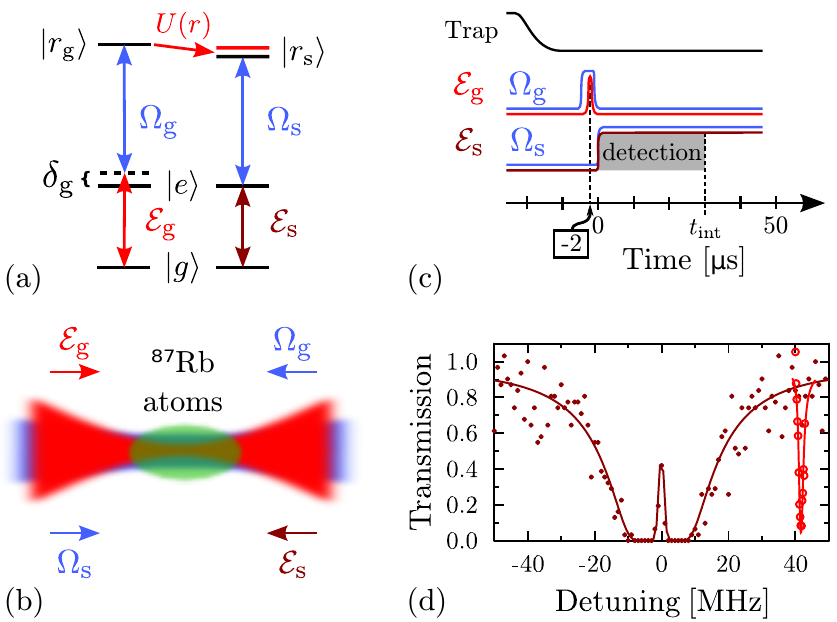}
\caption{\label{setup}
(a) Level scheme, (b) simplified schematic, and (c) pulse sequence of our all-optical transistor. (d) The absorption spectrum for the source field (dots) over the full intermediate state absorption valley shows the EIT window on resonance; the gate field spectrum (circles) is taken around the two-photon resonance at $\delta_g = \unit[40]{\mega\hertz}$. The solid lines are fits to the EIT spectra. The linewidth of the source EIT window $\Delta\nu = \unit[2]{\mega\hertz}$ and the optical depth $\text{OD}=25$ of our system are extracted from the brown curve.}
\end{center}
\end{figure}

The level scheme and geometry of our transistor are illustrated in Fig.~\ref{setup}~(a,b). Photons in the weak gate field $\mathcal{E}_\text{g}$ are stored as Rydberg excitations in an atomic ensemble by coupling the ground state $\Ket{g}$ to the Rydberg state $\Ket{r_\text{g}}$ via the strong control field $\Omega_\text{g}$ with an intermediate state detuning $\delta_\text{g}$ in a two-photon Raman process. After this storage process, the source field $\mathcal{E}_\text{s}$ is sent through the medium at reduced velocity due to resonant EIT provided by the control field $\Omega_\text{s}$ coupling to the Rydberg state $\Ket{r_\text{s}}$. In the absence of a gate excitation and at a low source photon rate, individual source photons travel through the transparent medium.  The strong van-der-Waals interaction between the Rydberg states $\Ket{r_\text{g}}$ and $\Ket{r_\text{s}}$ lifts the EIT condition within the inter-state blockade radius $r_\text{gs}$ around a gate excitation. In this blockaded volume, the atomic level structure reduces to an absorbing two-level system, i.e., transmission of source photons is attenuated by scattering from the intermediate state \cite{Lukin2011}. To observe this conditional switching, we record the number of transmitted source photons in a time interval $t_\text{int}$ after the gate excitation pulse, cf.\ Fig.~\ref{setup}~(c).

For the experimental realization of this scheme, we prepare $2.5 \cdot 10^{4}$ $^{87}$Rb atoms pumped into the $\Ket{g}=\Ket{5\text{S}_{1/2},F = 2, m_F = 2}$ state in an optical dipole trap (ODT). The final temperature of the ensemble is $ T = \unit[40]{\micro\kelvin}$ and the radial and axial radii of the cloud are $\sigma_\text{r} = \unit[25]{\micro\metre} $ and $\sigma_\text{l} = \unit[40]{\micro\metre}$. Originating from two lasers at $\sim \unit[780]{\nano\metre}$, the gate and source beams are tightly focussed ($1/e^2$ radius $\unit[5]{\micro\metre}$) into the medium along the long axis of the cloud from opposite directions. The overlapped beams are separated by polarization optics and we employ fiber-coupled avalanche photodetectors to resolve single photons (overall photon detection efficiency $0.31$). The two control beams at $\sim \unit[480]{\nano\metre}$ are sent through the medium counter-propagating to the respective gate and source beams. Photons in the gate pulse of duration $\unit[0.5]{\micro\second}$ (FWHM), blue-detuned by $\delta_\text{g} = \unit[40]{\mega\hertz}$ from the $\Ket{g} \rightarrow  \Ket{e}=\Ket{5\text{P}_{3/2}, F = 3, m_F = 3 }$ transition, are stored in the $\Ket{r_\text{g}}=\Ket{90\text{S}_{1/2},m_J=1/2}$ state. The source field, resonantly ($\delta_s = 0$) coupled to the $\Ket{r_\text{s}}=\Ket{89\text{S}_{1/2},m_J=1/2}$ state, is turned on $\unit[2]{\micro\second}$ after the gate pulse. We execute the pulse sequence for Rydberg excitation directly after switching off the ODT, to avoid level shifts due to the AC-Stark effect of the trap light.

Fig.~\ref{setup}~(d) shows measurements of the gate two-photon spectral line at $\delta_\text{g} = \unit[40]{\mega\hertz}$ (red circles) and the EIT window for the source photons (brown dots) centered in the broad absorption valley of the $\Ket{g} \leftrightarrow  \Ket{e}$ transition. Both lines show a FWHM linewidth of $\Delta\nu_\text{EIT} = \Delta\nu_\text{gate} = \unit[2]{\mega\hertz}$. The total optical depth of our system $\text{OD} = 25$ is extracted from a fit to the EIT spectrum.

For our choice of Rydberg states $\Ket{r_\text{s}}$ and $\Ket{r_\text{g}}$,  the inter-state interaction energy becomes larger than $\Delta \nu_\text{EIT}$ at distances $r_\text{gg} \approx r_\text{ss} \approx r_\text{gs} \approx \unit[15]{\micro\metre}$, the blockade radii of the state combinations involved in our scheme.  As $r_\text{gs}$ is significantly larger than the radial extent of the gate and source beams, the essential requirement for an efficient transistor is met.

%
\begin{figure}[!t]
\begin{center}
\input{fig/contrast.inc}
\caption{\label{contrast} Switch contrast (red) as function of (a) mean number of incident gate photons and (b) mean number of stored photons according to eq.~(\ref{eq:stored}). The dashed line indicates the fundamental limit set by the photon statistics of the coherent gate input. Black data points represent the calculated switch contrast expected for (a) one-, two- and three-photon Fock input states or (b) deterministic single and two stored gate excitations.}
\end{center}
\end{figure}
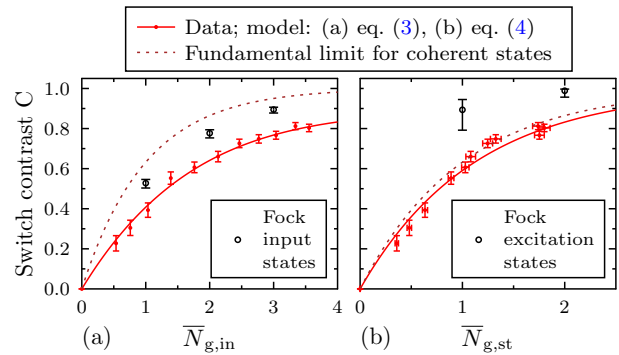
We first investigate the transistor performance for a weak source input photon rate $R_\text{s,in} = \unit[0.69(1)]{\micro\second^{-1}}$. In Fig.~\ref{contrast}~(a) we plot the switch contrast in the source beam transmission as a function of the mean incoming gate photon number $\overline{N}_\text{g,in}$. Source photon detection spans an integration time of $t_\text{int}=\unit[30]{\micro\second}$ to acquire significant statistical data, yet $t_\text{int}$ is small compared to the temperature limited retention time of the gate excitation in the source beam volume. For an average gate photon number of $\overline{N}_\text{g,in} = 1.04(3) $, we observe a switch contrast $C_\text{coh} = 0.39(4)$. The switch contrast is mainly determined by the Poissonian statistics of our coherent gate photons. Given $\overline{N}_\text{g,in}$, the probability to have zero photons in such a gate pulse is $p(0) = e^{-\overline{N}_\text{g,in}}$, which sets a fundamental upper bound to the switch contrast, in other words, a perfect switch with coherent gate photons has a switch contrast $C_\text{coh} = 1- e^{-\overline{N}_\text{g,in}}$ (dashed line in Fig.~\ref{contrast}~(a). How close our switch approaches this fundamental limit depends on the gate photon storage efficiency and the source attenuation caused by a single gate excitation. To estimate the storage efficiency, we monitor the gate pulse transmission $\overline{N}_\text{g,out}$ and take into account the finite absorption $A_\text{ge} =0.15$ of gate photons from the intermediate level $\Ket{e}$ (measured independently at $\Omega_\text{g}=0$). The mean number of stored gate photons is given by
\begin{equation}
\overline{N}_\text{g,st}= (1-A_\text{ge}) \cdot  \overline{N}_\text{g,in} - \overline{N}_\text{g,out} \text{ .}
\label{eq:stored}
\end{equation}
In Fig.~\ref{contrast}~(b) we plot the switch contrast versus $\overline{N}_\text{g,st}$. We note that this estimate is an upper limit, as there may be mechanisms removing gate excitations before or during the switch operation but not resulting in a gate photon count. These processes cannot be significant, though, as our corrected switch contrast measurement in Fig.~\ref{contrast}~(b) would then become better than the fundamental limit for gate excitations following Poissonian statistics.

To extract the mean switch contrast caused by a single incoming gate photon Fock state, we take into account the photon statistics of the gate pulses and the saturation of gate photon storage due to Rydberg blockade. The calculated gate-gate blockade radius $r_\text{gg} \approx \unit[15] {\micro\metre}$ suggests that it becomes very unlikely to store more than 3 gate photons simultaneously, which is in agreement with the observed storage efficiency for large $\overline{N}_\text{g,in}$. Assuming that a single incoming gate photon causes an attenuation of the source beam by $e^{-\text{OD}_\text{sp}}$, we write the expected switch contrast as
\begin{equation}\label{eq:Cs}
 C(\overline{N}_\text{g,in}) =1 -  \sum_{k=0}^{\infty} \frac{\overline{N}_\text{g,in}^k e^{-\overline{N}_\text{g,in}}}{k!}e^{-\text{min}(k,3)\cdot \text{OD}_\text{sp}}\text{ ,}
\end{equation}
where we account for the blockade by allowing all photon number states $k > 3$ to contribute only 3 gate excitations. This simple model is motivated by more sophisticated work on statistics of hard rods \cite{Lesanovsky2013} and fits our data very well. Applying the model from eq.~(\ref{eq:Cs}) to our data (solid line in Fig.~\ref{contrast}~a) we obtain $\text{OD}_\text{sp} = 0.75(5)$ caused by a single incoming gate photon. We stress that this value does not correct for any loss or imperfections of our switch. In analogy to eq.~(\ref{eq:Cs}), the switch contrast for stored gate excitations is modeled by
\begin{equation}\label{eq:Cstored}
 C(\overline{N}_\text{g,st}) =1 -  \sum_{k=0}^{\infty} \frac{\overline{N}_\text{g,st}^k e^{-\overline{N}_\text{g,st}}}{k!}e^{-\text{min}(k,3)\cdot \text{OD}_\text{st}}\text{ .}
\end{equation}
We determine a mean optical depth $\text{OD}_\text{st} = 2.2(7)$ per stored gate excitation. This value results from the elongated shape of our cloud, with gate excitations located in low density regions reducing the average $\text{OD}_\text{st}$  per excitation. Longitudinal compression to $\sigma_\text{l} = r_\text{gs}/2$ would increase $\text{OD}_\text{st}$ to the total optical depth of our system \cite{Duerr2014}.

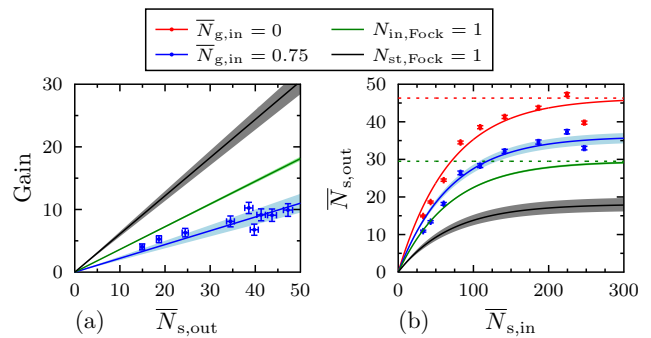
\begin{figure}[!t]
\begin{center}
\input{fig/transistor90us.inc}
\caption{\label{transistor} (a) Gain of our transistor, measured for coherent gate input $\overline{N}_\text{g,in}=0.75(3)$ (blue data) and extrapolated to single photon Fock state input (green line) and single stored excitation (black line). (b) Source photon transfer function without (red) and with coherent gate input $\overline{N}_\text{g,in}=0.75(3)$ (blue). We observe a constant switch contrast between the two data sets over the whole source input range. The green (black) solid line are again the estimated behavior of the system for a single-photon Fock input state (a single stored excitation). Shaded regions are error estimates.}
\end{center}
\end{figure}

Next, we investigate how many source photons can be switched by our system. To quantify the gate-induced change in source transmission, we consider the optical gain $G = \overline{N}_\text{s,out}^\text{no gate} - \overline{N}_\text{s,out}^\text{with gate}$ \cite{Vuletic2013}. In Fig.~\ref{transistor}~(a), we plot the measured gain for an average input of gate photons $\overline{N}_\text{g,in}=0.75(3)$ and $t_\text{int} = \unit[90]{\micro\second}$ integration time (blue data points). For this gate input, we observe a maximum gain $G(\overline{N}_\text{g,in}=0.75)=10(1)$.

Further increase of the gain at fixed gate input is limited by the self-blockade of the source beam, which results in nonlinear source transmission even in the absence of gate photons \cite{Adams2010,Vuletic2012}. The red (blue) data points in Fig.~\ref{transistor}~(b) show the source photon transfer function when $\overline{N}_\text{g,in}=0$ ($\overline{N}_\text{g,in}=0.75(3)$). The saturation of transmitted photons even in the absence of gate photons is well described by the expression $\overline{N}_\text{s,out}  = a \cdot (1-e^{-\overline{N}_\text{s,in} /b})$, where $a$ is the maximum number of photons transmitted through the medium and $b$ is the number of photons in the detection period where the self-nonlinear regime is reached\cite{Duerr2014}. Fitting this model to the data without gate input yields $a=46$ and $b=70$ (red line in Fig.~\ref{transistor}~(b)). Repeating the switch contrast analysis for low source input, we obtain $C_\text{coh,\unit[90]{\micro\second}} = 0.22(3)$ for the coherent gate input with $\overline{N}_\text{g,in}=0.75(3)$. Multiplying the self-nonlinearity fit  with $(1-C_\text{coh,\unit[90]{\micro\second}})$ results in the blue line in Fig.~\ref{transistor}~(b), which is in good agreement with the switched data points at all source photon inputs. From this we conclude that the stored gate excitations are unaffected by incoming source photon numbers as large as $\overline{N}_\text{s,in} \approx 250$, showing that our Rydberg-EIT transistor can achieve robust high gain $G\gg 1$ if the source photon loss due to self-blockade is overcome. One obvious improvement of our setup is to reduce the longitudinal extent of the cloud, ideally to $\sigma_\text{l} = r_\text{gs}/2$, while keeping the total OD constant. There is a fundamental limit on the peak atomic density set by Rydberg-ground state interaction, though \cite{Pfau2013,Duerr2014}. Our two-color approach also offers the option of tuning the inter-state interaction such that $r_\text{gs} \gg r_\text{ss}$ by exploiting Stark-tuned two-color F\"orster resonances \cite{vanLinden2008,Weidemueller2013b} or by directly working with dipole-coupled states.

Finally, the solid black (green) lines in Fig.~\ref{transistor}~(a,b) show the expected gain and source transfer function for a single photon Fock gate input (a deterministic single stored Rydberg excitation). The mean optical depth $\text{OD}_\text{sp,\unit[90]{\micro\second}} = 0.45(1)$ caused by a single photon Fock state gate input and $\text{OD}_\text{st,\unit[90]{\micro\second}} = 0.94(10)$ caused by a single stored excitation are obtained from fitting eqs.~(\ref{eq:Cs},\ref{eq:Cstored}) to this dataset. The reduction of mean optical depth stems from the fact that $t_\text{int} = \unit[90]{\micro\second}$ is similar to the average fly-away time of the stored excitations. For a single excitation, we calculate the maximally achievable gain of our current system as $G_\text{st} = 28(2)$.

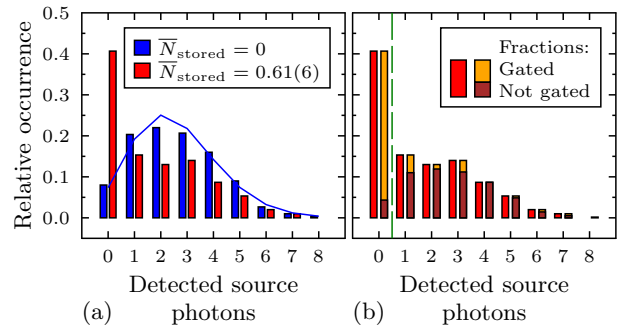
\begin{figure}[t]
\begin{center}
\input{fig/histostacked.inc}
\caption{\label{histo}~(a) Histograms of single shot detection events with and without stored gate excitation (red bars and blue bars respectively). The strong shift of the red histogram originates from the switch effect of single or two gate Rydberg excitations in the medium. (b) Accounting for the Poissonian statistics of gate Rydberg excitations, we decompose the red histogram in (a) into gated and not-gated contributions. The dashed green line is the discrimination level which predicts the presence of a Rydberg atom in a single shot with the highest fidelity of 0.72(4).}
\end{center}
\end{figure}

As a first application of our transistor we demonstrate single shot optical non-destructive detection of single Rydberg atoms in a dense background gas with high fidelity \cite{Weidemueller2012,Lesanovsky2011}. In Fig.~\ref{histo}~(a) we show histograms of registered source photon events (not corrected for our photon detection efficiency of $0.31$) obtained from many ($250$ each) individual runs of our transistor scheme for two different numbers of mean stored gate Rydberg excitations. For $\overline{N}_\text{stored} = 0$ the histogram (blue bars) is Poissonian (solid blue line) within statistical errors. In contrast, the histogram recorded for $\overline{N}_\text{stored} = 0.61(6) $ (red bars) shows a clearly non-Poissonian distribution with a strong redistribution towards zero detection events. Accounting again for the statistics of our classical gate pulses we can decompose the observed histogram into contributions with and without gate Rydberg excitations. The brown and orange histograms in Fig.~\ref{histo}~(b) show the contributions of gated and not gated experiments to the measured red histogram. The gated sub-ensemble consists mostly of runs with a single ($\unit[73]{\%}$) or two ($\unit[22]{\%}$) Rydberg excitation(s) in the cloud. A single gate excitation Fock state \cite{Walker2014} would cause a very similar histogram due to the high switch contrast. The dashed green line shows the discrimination level for the decision in a single shot whether one or more Rydberg atoms were present or not \cite{Rempe2010}. From the decomposed histograms we calculate the fidelity of this prediction for the dataset shown in Fig.~\ref{histo} to be $\mathcal{F}=0.72(4)$.

The limiting factor on the detection efficiency is again the self-blockade of the source photons, which prevents larger numbers of detection events for the non-gated case. Possible improvements are similar to the one listed above for the transistor, if no spatial resolution of the detection is desired. For spatially resolved imaging of Rydberg ensembles \cite{Weidemueller2012,Lesanovsky2011}, tuning of the inter-state interaction is restricted, as $r_\text{gs}$ must remain smaller than half the distance between gate Rydberg atoms, given by $\sim r_\text{gg}/2$ in a blockaded ensemble, to prevent spatial blurring.

In summary, we have demonstrated a free-space single-photon transistor based on two-color Rydberg interaction. Further improvements of our system could enable a high-gain, high efficiency optical transistor, so far only realized in a cavity QED setup \cite{Vuletic2013}. One key issue we will address in the future is the retrieval of gate photon(s) after the switch process \cite{Kuzmich2012b,Duerr2014}, which could enable multi-photon entanglement protocols and creation of non-classical light-states with large photon numbers \cite{Schoelkopf2013b}. We have implemented optical non-destructive Rydberg atom detection at the single particle level, which can be extended to spatially resolved imaging of Rydberg ensembles \cite{Weidemueller2012,Lesanovsky2011}. Increasing the gate photon storage efficiency would also enable high fidelity single photon detection \cite{Buechler2011}, retrieval of the gate photon would make this scheme non-destructive \cite{Rempe2013}. Finally, our system is a highly sensitive probe for studying Rydberg interaction on the few-particle level \cite{Browaeys2013,Lesanovsky2014}. In particular, the combination of two independently controlled Rydberg-EIT schemes enables novel fields of study, such as two-color F\"{o}rster resonances \cite{vanLinden2008,Weidemueller2013b} or a two-photon phase gate based on Rydberg-polariton collision \cite{Lukin2011,Xiao2014}.

Related experiments have been performed in the group of G.\ Rempe at the Max Planck Institute of Quantum Optics\cite{Rempe2014}.

We thank S.\ Weber, T.\ Pfau, I.\ Lesanovsky, W.\ Li, D.\ Viscor for fruitful discussions on various aspects of our experiments. This work is funded by the German Research Foundation through Emmy-Noether-grant HO 4787/1-1 and within the SFB/TRR21. H.G.\ acknowledges support from the Carl-Zeiss Foundation.

\bibliographystyle{apsrev4-1}
\bibliography{biblio}

\end{document}

%% file: fig/contrast.inc
\setlength{\unitlength}{1cm}%
\noindent{}\begin{picture}(8.575,4.825)(0,0)%
\put(0,0){\includegraphics{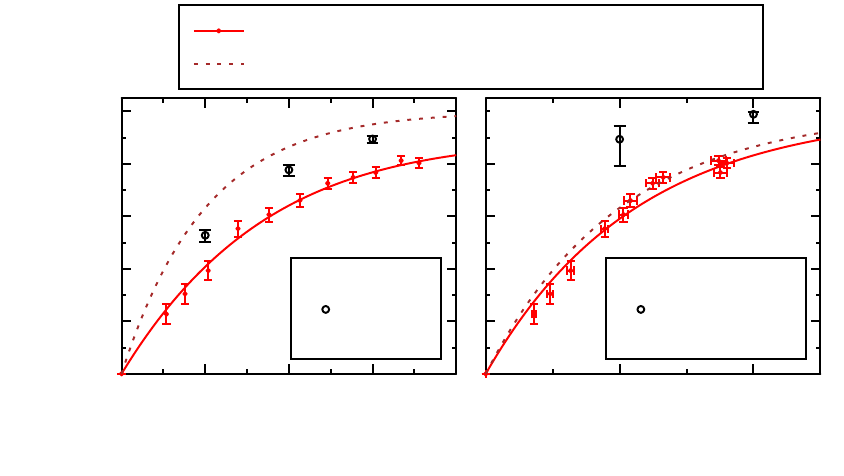}}
\put(0.735349,0.955109){\makebox(0,0)[lb]{\scalebox{0.891772}{{\footnotesize 0.0}}}}
\put(0.735349,1.48844){\makebox(0,0)[lb]{\scalebox{0.891772}{{\footnotesize 0.2}}}}
\put(0.735349,2.02178){\makebox(0,0)[lb]{\scalebox{0.891772}{{\footnotesize 0.4}}}}
\put(0.735349,2.55511){\makebox(0,0)[lb]{\scalebox{0.891772}{{\footnotesize 0.6}}}}
\put(0.735349,3.08844){\makebox(0,0)[lb]{\scalebox{0.891772}{{\footnotesize 0.8}}}}
\put(0.735349,3.62178){\makebox(0,0)[lb]{\scalebox{0.891772}{{\footnotesize 1.0}}}}
\put(0.577192,1.19688){\rotatebox{90}{\makebox(0,0)[lb]{\scalebox{0.899595}{{\normalsize Switch contrast C}}}}}
\put(1.16777,0.716783){\makebox(0,0)[lb]{\scalebox{0.891772}{{\footnotesize 0}}}}
\put(2.01777,0.716783){\makebox(0,0)[lb]{\scalebox{0.891772}{{\footnotesize 1}}}}
\put(2.86777,0.716783){\makebox(0,0)[lb]{\scalebox{0.891772}{{\footnotesize 2}}}}
\put(3.71777,0.716783){\makebox(0,0)[lb]{\scalebox{0.891772}{{\footnotesize 3}}}}
\put(4.56777,0.716783){\makebox(0,0)[lb]{\scalebox{0.891772}{{\footnotesize 4}}}}
\put(2.57874,0.187318){\makebox(0,0)[lb]{\scalebox{0.899595}{{\normalsize $\overline{N}_\text{g,in}$}}}}
\put(1.23528,0.218822){\makebox(0,0)[lb]{\scalebox{0.899595}{{\normalsize (a)}}}}
\put(3.64357,1.90858){\makebox(0,0)[lb]{\scalebox{0.997053}{{\scriptsize Fock}}}}
\put(3.64357,1.56694){\makebox(0,0)[lb]{\scalebox{0.997053}{{\scriptsize input}}}}
\put(3.64357,1.31813){\makebox(0,0)[lb]{\scalebox{0.997053}{{\scriptsize states}}}}
\put(4.7729,1.03528){\makebox(0,0)[lb]{\scalebox{0.891772}{{\footnotesize }}}}
\put(4.7729,1.56861){\makebox(0,0)[lb]{\scalebox{0.891772}{{\footnotesize }}}}
\put(4.7729,2.10194){\makebox(0,0)[lb]{\scalebox{0.891772}{{\footnotesize }}}}
\put(4.86777,0.716783){\makebox(0,0)[lb]{\scalebox{0.891772}{{\footnotesize 0}}}}
\put(6.22777,0.716783){\makebox(0,0)[lb]{\scalebox{0.891772}{{\footnotesize 1}}}}
\put(7.58777,0.716783){\makebox(0,0)[lb]{\scalebox{0.891772}{{\footnotesize 2}}}}
\put(6.28296,0.187318){\makebox(0,0)[lb]{\scalebox{0.899595}{{\normalsize $\overline{N}_\text{g,st}$}}}}
\put(4.93528,0.218822){\makebox(0,0)[lb]{\scalebox{0.899595}{{\normalsize (b)}}}}
\put(6.84567,1.90858){\makebox(0,0)[lb]{\scalebox{0.997053}{{\scriptsize Fock}}}}
\put(6.84567,1.61336){\makebox(0,0)[lb]{\scalebox{0.997053}{{\scriptsize excitation}}}}
\put(6.84567,1.31813){\makebox(0,0)[lb]{\scalebox{0.997053}{{\scriptsize states}}}}
\put(2.60497,4.35497){\makebox(0,0)[lb]{\scalebox{0.990858}{{\footnotesize Data; model: (a) eq.~(\ref{eq:Cs}), (b) eq.~(\ref{eq:Cstored})}}}}
\put(2.60497,4.08508){\makebox(0,0)[lb]{\scalebox{0.990858}{{\footnotesize Fundamental limit for coherent states}}}}
\end{picture}%

%% file: fig/transistor90us.inc
\setlength{\unitlength}{1cm}%
\noindent{}\begin{picture}(8.575,4.575)(0,0)%
\put(0,0){\includegraphics{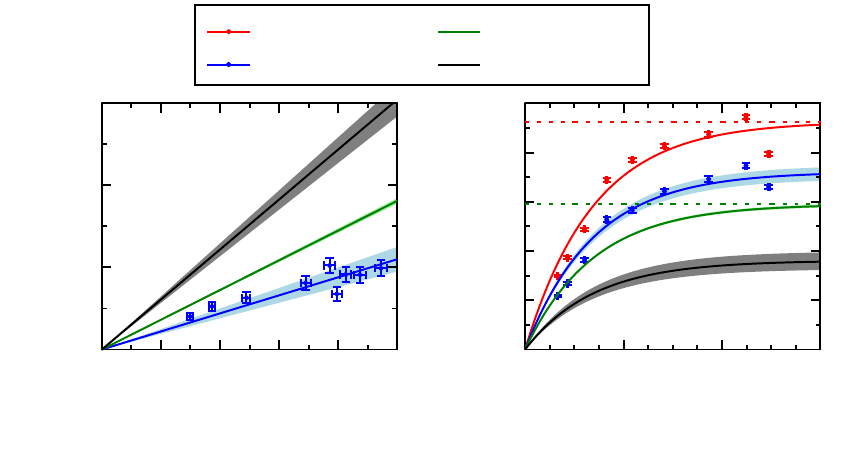}}
\put(0.7421,0.955109){\makebox(0,0)[lb]{\scalebox{0.891772}{{\footnotesize 0}}}}
\put(0.607079,1.78844){\makebox(0,0)[lb]{\scalebox{0.891772}{{\footnotesize 10}}}}
\put(0.607079,2.62178){\makebox(0,0)[lb]{\scalebox{0.891772}{{\footnotesize 20}}}}
\put(0.607079,3.45511){\makebox(0,0)[lb]{\scalebox{0.891772}{{\footnotesize 30}}}}
\put(0.448922,1.94983){\rotatebox{90}{\makebox(0,0)[lb]{\scalebox{0.899595}{{\normalsize Gain}}}}}
\put(0.967767,0.716783){\makebox(0,0)[lb]{\scalebox{0.891772}{{\footnotesize 0}}}}
\put(1.50026,0.716783){\makebox(0,0)[lb]{\scalebox{0.891772}{{\footnotesize 10}}}}
\put(2.10026,0.716783){\makebox(0,0)[lb]{\scalebox{0.891772}{{\footnotesize 20}}}}
\put(2.70026,0.716783){\makebox(0,0)[lb]{\scalebox{0.891772}{{\footnotesize 30}}}}
\put(3.30026,0.716783){\makebox(0,0)[lb]{\scalebox{0.891772}{{\footnotesize 40}}}}
\put(3.90026,0.716783){\makebox(0,0)[lb]{\scalebox{0.891772}{{\footnotesize 50}}}}
\put(2.11334,0.187318){\makebox(0,0)[lb]{\scalebox{0.899595}{{\normalsize $\overline{N}_\text{s,out}$}}}}
\put(1.03528,0.218822){\makebox(0,0)[lb]{\scalebox{0.899595}{{\normalsize (a)}}}}
\put(5.0421,0.955109){\makebox(0,0)[lb]{\scalebox{0.891772}{{\footnotesize 0}}}}
\put(4.90708,1.45511){\makebox(0,0)[lb]{\scalebox{0.891772}{{\footnotesize 10}}}}
\put(4.90708,1.95511){\makebox(0,0)[lb]{\scalebox{0.891772}{{\footnotesize 20}}}}
\put(4.90708,2.45511){\makebox(0,0)[lb]{\scalebox{0.891772}{{\footnotesize 30}}}}
\put(4.90708,2.95511){\makebox(0,0)[lb]{\scalebox{0.891772}{{\footnotesize 40}}}}
\put(4.90708,3.45511){\makebox(0,0)[lb]{\scalebox{0.891772}{{\footnotesize 50}}}}
\put(4.74892,1.86334){\rotatebox{90}{\makebox(0,0)[lb]{\scalebox{0.899595}{{\normalsize $\overline{N}_\text{s,out}$}}}}}
\put(5.26777,0.716783){\makebox(0,0)[lb]{\scalebox{0.891772}{{\footnotesize 0}}}}
\put(6.13486,0.716783){\makebox(0,0)[lb]{\scalebox{0.891772}{{\footnotesize 100}}}}
\put(7.13486,0.716783){\makebox(0,0)[lb]{\scalebox{0.891772}{{\footnotesize 200}}}}
\put(8.13486,0.716783){\makebox(0,0)[lb]{\scalebox{0.891772}{{\footnotesize 300}}}}
\put(6.4914,0.187318){\makebox(0,0)[lb]{\scalebox{0.899595}{{\normalsize $\overline{N}_\text{s,in}$}}}}
\put(5.33528,0.218822){\makebox(0,0)[lb]{\scalebox{0.899595}{{\normalsize (b)}}}}
\put(2.65005,4.11454){\makebox(0,0)[lb]{\scalebox{0.972648}{{\scriptsize $\overline{N}_\text{g,in}=0$}}}}
\put(2.65005,3.78078){\makebox(0,0)[lb]{\scalebox{0.972648}{{\scriptsize $\overline{N}_\text{g,in}=0.75$}}}}
\put(4.9901,4.11454){\makebox(0,0)[lb]{\scalebox{0.972648}{{\scriptsize $N_\text{in,Fock}=1$}}}}
\put(4.9901,3.78078){\makebox(0,0)[lb]{\scalebox{0.972648}{{\scriptsize $N_\text{st,Fock}=1$}}}}
\end{picture}%

%% file: fig/histostacked.inc
\setlength{\unitlength}{1cm}%
\noindent{}\begin{picture}(8.575,4.575)(0,0)%
\put(0,0){\includegraphics{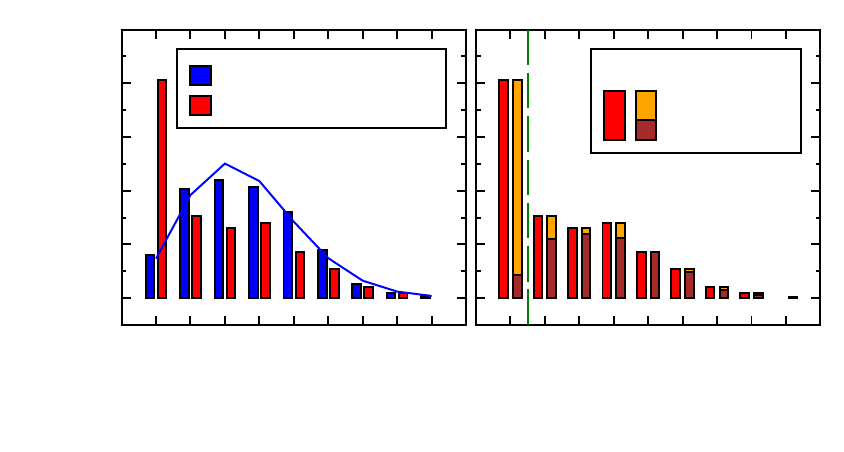}}
\put(0.714252,1.47784){\makebox(0,0)[lb]{{\scriptsize 0.0}}}
\put(0.714252,2.02329){\makebox(0,0)[lb]{{\scriptsize 0.1}}}
\put(0.714252,2.56875){\makebox(0,0)[lb]{{\scriptsize 0.2}}}
\put(0.714252,3.1142){\makebox(0,0)[lb]{{\scriptsize 0.3}}}
\put(0.714252,3.65965){\makebox(0,0)[lb]{{\scriptsize 0.4}}}
\put(0.714252,4.20511){\makebox(0,0)[lb]{{\scriptsize 0.5}}}
\put(0.556095,1.3992){\rotatebox{90}{\makebox(0,0)[lb]{{\small Relative occurrence}}}}
\put(1.51355,0.966783){\makebox(0,0)[lb]{{\scriptsize 0}}}
\put(1.86355,0.966783){\makebox(0,0)[lb]{{\scriptsize 1}}}
\put(2.21355,0.966783){\makebox(0,0)[lb]{{\scriptsize 2}}}
\put(2.56355,0.966783){\makebox(0,0)[lb]{{\scriptsize 3}}}
\put(2.91355,0.966783){\makebox(0,0)[lb]{{\scriptsize 4}}}
\put(3.26355,0.966783){\makebox(0,0)[lb]{{\scriptsize 5}}}
\put(3.61355,0.966783){\makebox(0,0)[lb]{{\scriptsize 6}}}
\put(3.96355,0.966783){\makebox(0,0)[lb]{{\scriptsize 7}}}
\put(4.31355,0.966783){\makebox(0,0)[lb]{{\scriptsize 8}}}
\put(1.8587,0.14196){\makebox(0,0)[lb]{{\small \pbox{4cm}{Detected source\\ photons}}}}
\put(1.23528,0.135278){\makebox(0,0)[lb]{{\small (a)}}}
\put(2.25475,3.6866){\makebox(0,0)[lb]{{\scriptsize $\overline{N}_\text{stored} = 0$}}}
\put(2.25475,3.36086){\makebox(0,0)[lb]{{\scriptsize $\overline{N}_\text{stored} = 0.61(6)$}}}
\put(4.6729,1.55801){\makebox(0,0)[lb]{{\scriptsize }}}
\put(4.6729,2.10346){\makebox(0,0)[lb]{{\scriptsize }}}
\put(4.6729,2.64891){\makebox(0,0)[lb]{{\scriptsize }}}
\put(5.11355,0.966783){\makebox(0,0)[lb]{{\scriptsize 0}}}
\put(5.46355,0.966783){\makebox(0,0)[lb]{{\scriptsize 1}}}
\put(5.81355,0.966783){\makebox(0,0)[lb]{{\scriptsize 2}}}
\put(6.16355,0.966783){\makebox(0,0)[lb]{{\scriptsize 3}}}
\put(6.51355,0.966783){\makebox(0,0)[lb]{{\scriptsize 4}}}
\put(6.86355,0.966783){\makebox(0,0)[lb]{{\scriptsize 5}}}
\put(7.21355,0.966783){\makebox(0,0)[lb]{{\scriptsize 6}}}
\put(7.56355,0.966783){\makebox(0,0)[lb]{{\scriptsize 7}}}
\put(7.91355,0.966783){\makebox(0,0)[lb]{{\scriptsize 8}}}
\put(5.4587,0.14196){\makebox(0,0)[lb]{{\small \pbox{4cm}{Detected source\\ photons}}}}
\put(4.83528,0.135278){\makebox(0,0)[lb]{{\small (b)}}}
\put(6.7788,3.75758){\makebox(0,0)[lb]{{\scriptsize Fractions:}}}
\put(6.7788,3.46235){\makebox(0,0)[lb]{{\scriptsize Gated}}}
\put(6.7788,3.12071){\makebox(0,0)[lb]{{\scriptsize Not gated}}}
\end{picture}%

%% file: Single_Photon_Transistor_Mediated_by_Inter-State_Rydberg_Interaction.bbl
\begin{thebibliography}{40}%
\makeatletter
\providecommand \@ifxundefined [1]{%
 \@ifx{#1\undefined}
}%
\providecommand \@ifnum [1]{%
 \ifnum #1\expandafter \@firstoftwo
 \else \expandafter \@secondoftwo
 \fi
}%
\providecommand \@ifx [1]{%
 \ifx #1\expandafter \@firstoftwo
 \else \expandafter \@secondoftwo
 \fi
}%
\providecommand \natexlab [1]{#1}%
\providecommand \enquote  [1]{``#1''}%
\providecommand \bibnamefont  [1]{#1}%
\providecommand \bibfnamefont [1]{#1}%
\providecommand \citenamefont [1]{#1}%
\providecommand \href@noop [0]{\@secondoftwo}%
\providecommand \href [0]{\begingroup \@sanitize@url \@href}%
\providecommand \@href[1]{\@@startlink{#1}\@@href}%
\providecommand \@@href[1]{\endgroup#1\@@endlink}%
\providecommand \@sanitize@url [0]{\catcode `\\12\catcode `\$12\catcode
  `\&12\catcode `\#12\catcode `\^12\catcode `\_12\catcode `\%12\relax}%
\providecommand \@@startlink[1]{}%
\providecommand \@@endlink[0]{}%
\providecommand \url  [0]{\begingroup\@sanitize@url \@url }%
\providecommand \@url [1]{\endgroup\@href {#1}{\urlprefix }}%
\providecommand \urlprefix  [0]{URL }%
\providecommand \Eprint [0]{\href }%
\providecommand \doibase [0]{http://dx.doi.org/}%
\providecommand \selectlanguage [0]{\@gobble}%
\providecommand \bibinfo  [0]{\@secondoftwo}%
\providecommand \bibfield  [0]{\@secondoftwo}%
\providecommand \translation [1]{[#1]}%
\providecommand \BibitemOpen [0]{}%
\providecommand \bibitemStop [0]{}%
\providecommand \bibitemNoStop [0]{.\EOS\space}%
\providecommand \EOS [0]{\spacefactor3000\relax}%
\providecommand \BibitemShut  [1]{\csname bibitem#1\endcsname}%
\let\auto@bib@innerbib\@empty
\bibitem [{\citenamefont {Caulfield}\ and\ \citenamefont
  {Dolev}(2010)}]{Dolev2010}%
  \BibitemOpen
  \bibfield  {author} {\bibinfo {author} {\bibfnamefont {H.~J.}\ \bibnamefont
  {Caulfield}}\ and\ \bibinfo {author} {\bibfnamefont {S.}~\bibnamefont
  {Dolev}},\ }\href {\doibase doi:10.1038/nphoton.2010.94} {\bibfield
  {journal} {\bibinfo  {journal} {Nature Photonics}\ }\textbf {\bibinfo
  {volume} {4}},\ \bibinfo {pages} {261} (\bibinfo {year} {2010})}\BibitemShut
  {NoStop}%
\bibitem [{\citenamefont {O'Brien}\ \emph {et~al.}(2009)\citenamefont
  {O'Brien}, \citenamefont {Furusawa},\ and\ \citenamefont
  {Vu\v{c}kovi\'c}}]{Vuckovic2009}%
  \BibitemOpen
  \bibfield  {author} {\bibinfo {author} {\bibfnamefont {J.~L.}\ \bibnamefont
  {O'Brien}}, \bibinfo {author} {\bibfnamefont {A.}~\bibnamefont {Furusawa}}, \
  and\ \bibinfo {author} {\bibfnamefont {J.}~\bibnamefont {Vu\v{c}kovi\'c}},\
  }\href {http://dx.doi.org/10.1038/nphoton.2009.229} {\bibfield  {journal}
  {\bibinfo  {journal} {Nature Photonics}\ }\textbf {\bibinfo {volume} {3}},\
  \bibinfo {pages} {687} (\bibinfo {year} {2009})}\BibitemShut {NoStop}%
\bibitem [{\citenamefont {Birnbaum}\ \emph {et~al.}(2005)\citenamefont
  {Birnbaum}, \citenamefont {Boca}, \citenamefont {Miller}, \citenamefont
  {Boozer}, \citenamefont {Northup},\ and\ \citenamefont
  {Kimble}}]{Kimble2005}%
  \BibitemOpen
  \bibfield  {author} {\bibinfo {author} {\bibfnamefont {K.~M.}\ \bibnamefont
  {Birnbaum}}, \bibinfo {author} {\bibfnamefont {A.}~\bibnamefont {Boca}},
  \bibinfo {author} {\bibfnamefont {R.}~\bibnamefont {Miller}}, \bibinfo
  {author} {\bibfnamefont {A.~D.}\ \bibnamefont {Boozer}}, \bibinfo {author}
  {\bibfnamefont {T.~E.}\ \bibnamefont {Northup}}, \ and\ \bibinfo {author}
  {\bibfnamefont {H.~J.}\ \bibnamefont {Kimble}},\ }\href {\doibase
  doi:10.1038/nature03804} {\bibfield  {journal} {\bibinfo  {journal} {Nature}\
  }\textbf {\bibinfo {volume} {436}},\ \bibinfo {pages} {87} (\bibinfo {year}
  {2005})}\BibitemShut {NoStop}%
\bibitem [{\citenamefont {Wilk}\ \emph {et~al.}(2007)\citenamefont {Wilk},
  \citenamefont {Webster}, \citenamefont {Kuhn},\ and\ \citenamefont
  {Rempe}}]{Rempe2007}%
  \BibitemOpen
  \bibfield  {author} {\bibinfo {author} {\bibfnamefont {T.}~\bibnamefont
  {Wilk}}, \bibinfo {author} {\bibfnamefont {S.~C.}\ \bibnamefont {Webster}},
  \bibinfo {author} {\bibfnamefont {A.}~\bibnamefont {Kuhn}}, \ and\ \bibinfo
  {author} {\bibfnamefont {G.}~\bibnamefont {Rempe}},\ }\href {\doibase
  10.1126/science.1143835} {\bibfield  {journal} {\bibinfo  {journal}
  {Science}\ }\textbf {\bibinfo {volume} {317}},\ \bibinfo {pages} {488 }
  (\bibinfo {year} {2007})}\BibitemShut {NoStop}%
\bibitem [{\citenamefont {Kubanek}\ \emph {et~al.}(2008)\citenamefont
  {Kubanek}, \citenamefont {Ourjoumtsev}, \citenamefont {Schuster},
  \citenamefont {Koch}, \citenamefont {Pinkse}, \citenamefont {Murr},\ and\
  \citenamefont {Rempe}}]{Rempe2008b}%
  \BibitemOpen
  \bibfield  {author} {\bibinfo {author} {\bibfnamefont {A.}~\bibnamefont
  {Kubanek}}, \bibinfo {author} {\bibfnamefont {A.}~\bibnamefont
  {Ourjoumtsev}}, \bibinfo {author} {\bibfnamefont {I.}~\bibnamefont
  {Schuster}}, \bibinfo {author} {\bibfnamefont {M.}~\bibnamefont {Koch}},
  \bibinfo {author} {\bibfnamefont {P.~W.~H.}\ \bibnamefont {Pinkse}}, \bibinfo
  {author} {\bibfnamefont {K.}~\bibnamefont {Murr}}, \ and\ \bibinfo {author}
  {\bibfnamefont {G.}~\bibnamefont {Rempe}},\ }\href {\doibase
  10.1103/PhysRevLett.101.203602} {\bibfield  {journal} {\bibinfo  {journal}
  {Phys. Rev. Lett.}\ }\textbf {\bibinfo {volume} {101}},\ \bibinfo {pages}
  {203602} (\bibinfo {year} {2008})}\BibitemShut {NoStop}%
\bibitem [{\citenamefont {Tanji-Suzuki}\ \emph {et~al.}(2011)\citenamefont
  {Tanji-Suzuki}, \citenamefont {Chen}, \citenamefont {Landig}, \citenamefont
  {Simon},\ and\ \citenamefont {Vuleti\'{c}}}]{Vuletic2011}%
  \BibitemOpen
  \bibfield  {author} {\bibinfo {author} {\bibfnamefont {H.}~\bibnamefont
  {Tanji-Suzuki}}, \bibinfo {author} {\bibfnamefont {W.}~\bibnamefont {Chen}},
  \bibinfo {author} {\bibfnamefont {R.}~\bibnamefont {Landig}}, \bibinfo
  {author} {\bibfnamefont {J.}~\bibnamefont {Simon}}, \ and\ \bibinfo {author}
  {\bibfnamefont {V.}~\bibnamefont {Vuleti\'{c}}},\ }\href {\doibase
  10.1126/science.1208066} {\bibfield  {journal} {\bibinfo  {journal}
  {Science}\ }\textbf {\bibinfo {volume} {333}},\ \bibinfo {pages} {1266}
  (\bibinfo {year} {2011})}\BibitemShut {NoStop}%
\bibitem [{\citenamefont {O'Shea}\ \emph {et~al.}(2013)\citenamefont {O'Shea},
  \citenamefont {Junge}, \citenamefont {Volz},\ and\ \citenamefont
  {Rauschenbeutel}}]{Rauschenbeutel2013}%
  \BibitemOpen
  \bibfield  {author} {\bibinfo {author} {\bibfnamefont {D.}~\bibnamefont
  {O'Shea}}, \bibinfo {author} {\bibfnamefont {C.}~\bibnamefont {Junge}},
  \bibinfo {author} {\bibfnamefont {J.}~\bibnamefont {Volz}}, \ and\ \bibinfo
  {author} {\bibfnamefont {A.}~\bibnamefont {Rauschenbeutel}},\ }\href
  {\doibase 10.1103/PhysRevLett.111.193601} {\bibfield  {journal} {\bibinfo
  {journal} {Phys. Rev. Lett.}\ }\textbf {\bibinfo {volume} {111}},\ \bibinfo
  {pages} {193601} (\bibinfo {year} {2013})}\BibitemShut {NoStop}%
\bibitem [{\citenamefont {Michler}\ \emph {et~al.}(2000)\citenamefont
  {Michler}, \citenamefont {Kiraz}, \citenamefont {Becher}, \citenamefont
  {Schoenfeld}, \citenamefont {Petroff}, \citenamefont {Zhang}, \citenamefont
  {Hu},\ and\ \citenamefont {Imamo\u{g}lu}}]{Imamoglu2000}%
  \BibitemOpen
  \bibfield  {author} {\bibinfo {author} {\bibfnamefont {P.}~\bibnamefont
  {Michler}}, \bibinfo {author} {\bibfnamefont {A.}~\bibnamefont {Kiraz}},
  \bibinfo {author} {\bibfnamefont {C.}~\bibnamefont {Becher}}, \bibinfo
  {author} {\bibfnamefont {W.~V.}\ \bibnamefont {Schoenfeld}}, \bibinfo
  {author} {\bibfnamefont {P.~M.}\ \bibnamefont {Petroff}}, \bibinfo {author}
  {\bibfnamefont {L.}~\bibnamefont {Zhang}}, \bibinfo {author} {\bibfnamefont
  {E.}~\bibnamefont {Hu}}, \ and\ \bibinfo {author} {\bibfnamefont
  {A.}~\bibnamefont {Imamo\u{g}lu}},\ }\href {\doibase
  10.1126/science.290.5500.2282} {\bibfield  {journal} {\bibinfo  {journal}
  {Science}\ }\textbf {\bibinfo {volume} {290}},\ \bibinfo {pages} {2282}
  (\bibinfo {year} {2000})}\BibitemShut {NoStop}%
\bibitem [{\citenamefont {Englund}\ \emph {et~al.}(2007)\citenamefont
  {Englund}, \citenamefont {Faraon}, \citenamefont {Fushman}, \citenamefont
  {Stoltz}, \citenamefont {Petroff},\ and\ \citenamefont
  {Vu\v{c}kovi\'c}}]{Vuckovic2007}%
  \BibitemOpen
  \bibfield  {author} {\bibinfo {author} {\bibfnamefont {D.}~\bibnamefont
  {Englund}}, \bibinfo {author} {\bibfnamefont {A.}~\bibnamefont {Faraon}},
  \bibinfo {author} {\bibfnamefont {I.}~\bibnamefont {Fushman}}, \bibinfo
  {author} {\bibfnamefont {N.}~\bibnamefont {Stoltz}}, \bibinfo {author}
  {\bibfnamefont {P.}~\bibnamefont {Petroff}}, \ and\ \bibinfo {author}
  {\bibfnamefont {J.}~\bibnamefont {Vu\v{c}kovi\'c}},\ }\href {\doibase
  10.1038/nature06234} {\bibfield  {journal} {\bibinfo  {journal} {Nature}\
  }\textbf {\bibinfo {volume} {450}},\ \bibinfo {pages} {857} (\bibinfo {year}
  {2007})}\BibitemShut {NoStop}%
\bibitem [{\citenamefont {Fushman}\ \emph {et~al.}(2008)\citenamefont
  {Fushman}, \citenamefont {Englund}, \citenamefont {Faraon}, \citenamefont
  {Stoltz}, \citenamefont {Petroff},\ and\ \citenamefont
  {Vu\v{c}kovi\'c}}]{Vuckovic2008}%
  \BibitemOpen
  \bibfield  {author} {\bibinfo {author} {\bibfnamefont {I.}~\bibnamefont
  {Fushman}}, \bibinfo {author} {\bibfnamefont {D.}~\bibnamefont {Englund}},
  \bibinfo {author} {\bibfnamefont {A.}~\bibnamefont {Faraon}}, \bibinfo
  {author} {\bibfnamefont {N.}~\bibnamefont {Stoltz}}, \bibinfo {author}
  {\bibfnamefont {P.}~\bibnamefont {Petroff}}, \ and\ \bibinfo {author}
  {\bibfnamefont {J.}~\bibnamefont {Vu\v{c}kovi\'c}},\ }\href {\doibase
  10.1126/science.1154643} {\bibfield  {journal} {\bibinfo  {journal}
  {Science}\ }\textbf {\bibinfo {volume} {320}},\ \bibinfo {pages} {769}
  (\bibinfo {year} {2008})}\BibitemShut {NoStop}%
\bibitem [{\citenamefont {Volz}\ \emph {et~al.}(2012)\citenamefont {Volz},
  \citenamefont {Reinhard}, \citenamefont {Winger}, \citenamefont {Badolato},
  \citenamefont {Hennessy}, \citenamefont {Hu},\ and\ \citenamefont
  {Imamo\u{g}lu}}]{Imamoglu2012}%
  \BibitemOpen
  \bibfield  {author} {\bibinfo {author} {\bibfnamefont {T.}~\bibnamefont
  {Volz}}, \bibinfo {author} {\bibfnamefont {A.}~\bibnamefont {Reinhard}},
  \bibinfo {author} {\bibfnamefont {M.}~\bibnamefont {Winger}}, \bibinfo
  {author} {\bibfnamefont {A.}~\bibnamefont {Badolato}}, \bibinfo {author}
  {\bibfnamefont {K.~J.}\ \bibnamefont {Hennessy}}, \bibinfo {author}
  {\bibfnamefont {E.~L.}\ \bibnamefont {Hu}}, \ and\ \bibinfo {author}
  {\bibfnamefont {A.}~\bibnamefont {Imamo\u{g}lu}},\ }\href {\doibase
  doi:10.1038/nphoton.2012.181} {\bibfield  {journal} {\bibinfo  {journal}
  {Nature Photonics}\ }\textbf {\bibinfo {volume} {6}},\ \bibinfo {pages} {605}
  (\bibinfo {year} {2012})}\BibitemShut {NoStop}%
\bibitem [{\citenamefont {Chen}\ \emph {et~al.}(2013)\citenamefont {Chen},
  \citenamefont {Beck}, \citenamefont {B\"ucker}, \citenamefont {Gullans},
  \citenamefont {Lukin}, \citenamefont {Tanji-Suzuki},\ and\ \citenamefont
  {Vuleti\'{c}}}]{Vuletic2013}%
  \BibitemOpen
  \bibfield  {author} {\bibinfo {author} {\bibfnamefont {W.}~\bibnamefont
  {Chen}}, \bibinfo {author} {\bibfnamefont {K.~M.}\ \bibnamefont {Beck}},
  \bibinfo {author} {\bibfnamefont {R.}~\bibnamefont {B\"ucker}}, \bibinfo
  {author} {\bibfnamefont {M.}~\bibnamefont {Gullans}}, \bibinfo {author}
  {\bibfnamefont {M.~D.}\ \bibnamefont {Lukin}}, \bibinfo {author}
  {\bibfnamefont {H.}~\bibnamefont {Tanji-Suzuki}}, \ and\ \bibinfo {author}
  {\bibfnamefont {V.}~\bibnamefont {Vuleti\'{c}}},\ }\href {\doibase
  10.1126/science.1238169} {\bibfield  {journal} {\bibinfo  {journal}
  {Science}\ }\textbf {\bibinfo {volume} {341}},\ \bibinfo {pages} {768}
  (\bibinfo {year} {2013})}\BibitemShut {NoStop}%
\bibitem [{\citenamefont {Hwang}\ \emph {et~al.}(2009)\citenamefont {Hwang},
  \citenamefont {Pototschnig}, \citenamefont {Lettow}, \citenamefont {Zumofen},
  \citenamefont {Renn}, \citenamefont {Goetzinger},\ and\ \citenamefont
  {Sandoghdar}}]{Sandoghdar2009}%
  \BibitemOpen
  \bibfield  {author} {\bibinfo {author} {\bibfnamefont {J.}~\bibnamefont
  {Hwang}}, \bibinfo {author} {\bibfnamefont {M.}~\bibnamefont {Pototschnig}},
  \bibinfo {author} {\bibfnamefont {R.}~\bibnamefont {Lettow}}, \bibinfo
  {author} {\bibfnamefont {G.}~\bibnamefont {Zumofen}}, \bibinfo {author}
  {\bibfnamefont {A.}~\bibnamefont {Renn}}, \bibinfo {author} {\bibfnamefont
  {S.}~\bibnamefont {Goetzinger}}, \ and\ \bibinfo {author} {\bibfnamefont
  {V.}~\bibnamefont {Sandoghdar}},\ }\href {\doibase doi:10.1038/nature08134}
  {\bibfield  {journal} {\bibinfo  {journal} {Nature}\ }\textbf {\bibinfo
  {volume} {460}},\ \bibinfo {pages} {76} (\bibinfo {year} {2009})}\BibitemShut
  {NoStop}%
\bibitem [{\citenamefont {Bajcsy}\ \emph {et~al.}(2009)\citenamefont {Bajcsy},
  \citenamefont {Hofferberth}, \citenamefont {Bali\'c}, \citenamefont
  {Peyronel}, \citenamefont {Hafezi}, \citenamefont {Zibrov}, \citenamefont
  {Vuleti\'{c}},\ and\ \citenamefont {Lukin}}]{Lukin2009}%
  \BibitemOpen
  \bibfield  {author} {\bibinfo {author} {\bibfnamefont {M.}~\bibnamefont
  {Bajcsy}}, \bibinfo {author} {\bibfnamefont {S.}~\bibnamefont {Hofferberth}},
  \bibinfo {author} {\bibfnamefont {V.}~\bibnamefont {Bali\'c}}, \bibinfo
  {author} {\bibfnamefont {T.}~\bibnamefont {Peyronel}}, \bibinfo {author}
  {\bibfnamefont {M.}~\bibnamefont {Hafezi}}, \bibinfo {author} {\bibfnamefont
  {A.~S.}\ \bibnamefont {Zibrov}}, \bibinfo {author} {\bibfnamefont
  {V.}~\bibnamefont {Vuleti\'{c}}}, \ and\ \bibinfo {author} {\bibfnamefont
  {M.~D.}\ \bibnamefont {Lukin}},\ }\href {\doibase
  10.1103/PhysRevLett.102.203902} {\bibfield  {journal} {\bibinfo  {journal}
  {Phys. Rev. Lett.}\ }\textbf {\bibinfo {volume} {102}},\ \bibinfo {pages}
  {203902} (\bibinfo {year} {2009})}\BibitemShut {NoStop}%
\bibitem [{\citenamefont {Venkataraman}\ \emph {et~al.}(2011)\citenamefont
  {Venkataraman}, \citenamefont {Saha}, \citenamefont {Londero},\ and\
  \citenamefont {Gaeta}}]{Gaeta2011}%
  \BibitemOpen
  \bibfield  {author} {\bibinfo {author} {\bibfnamefont {V.}~\bibnamefont
  {Venkataraman}}, \bibinfo {author} {\bibfnamefont {K.}~\bibnamefont {Saha}},
  \bibinfo {author} {\bibfnamefont {P.}~\bibnamefont {Londero}}, \ and\
  \bibinfo {author} {\bibfnamefont {A.~L.}\ \bibnamefont {Gaeta}},\ }\href
  {\doibase 10.1103/PhysRevLett.107.193902} {\bibfield  {journal} {\bibinfo
  {journal} {Phys. Rev. Lett.}\ }\textbf {\bibinfo {volume} {107}},\ \bibinfo
  {pages} {193902} (\bibinfo {year} {2011})}\BibitemShut {NoStop}%
\bibitem [{\citenamefont {Vetsch}\ \emph {et~al.}(2010)\citenamefont {Vetsch},
  \citenamefont {Reitz}, \citenamefont {Sagu\'e}, \citenamefont {Schmidt},
  \citenamefont {Dawkins},\ and\ \citenamefont
  {Rauschenbeutel}}]{Rauschenbeutel2010}%
  \BibitemOpen
  \bibfield  {author} {\bibinfo {author} {\bibfnamefont {E.}~\bibnamefont
  {Vetsch}}, \bibinfo {author} {\bibfnamefont {D.}~\bibnamefont {Reitz}},
  \bibinfo {author} {\bibfnamefont {G.}~\bibnamefont {Sagu\'e}}, \bibinfo
  {author} {\bibfnamefont {R.}~\bibnamefont {Schmidt}}, \bibinfo {author}
  {\bibfnamefont {S.~T.}\ \bibnamefont {Dawkins}}, \ and\ \bibinfo {author}
  {\bibfnamefont {A.}~\bibnamefont {Rauschenbeutel}},\ }\href {\doibase
  10.1103/PhysRevLett.104.203603} {\bibfield  {journal} {\bibinfo  {journal}
  {Phys. Rev. Lett.}\ }\textbf {\bibinfo {volume} {104}},\ \bibinfo {pages}
  {203603} (\bibinfo {year} {2010})}\BibitemShut {NoStop}%
\bibitem [{\citenamefont {Saffman}\ \emph {et~al.}(2010)\citenamefont
  {Saffman}, \citenamefont {Walker},\ and\ \citenamefont
  {M\o{}lmer}}]{Saffman2010}%
  \BibitemOpen
  \bibfield  {author} {\bibinfo {author} {\bibfnamefont {M.}~\bibnamefont
  {Saffman}}, \bibinfo {author} {\bibfnamefont {T.~G.}\ \bibnamefont {Walker}},
  \ and\ \bibinfo {author} {\bibfnamefont {K.}~\bibnamefont {M\o{}lmer}},\
  }\href {\doibase 10.1103/RevModPhys.82.2313} {\bibfield  {journal} {\bibinfo
  {journal} {Rev. Mod. Phys.}\ }\textbf {\bibinfo {volume} {82}},\ \bibinfo
  {pages} {2313} (\bibinfo {year} {2010})}\BibitemShut {NoStop}%
\bibitem [{\citenamefont {Pritchard}\ \emph {et~al.}(2010)\citenamefont
  {Pritchard}, \citenamefont {Maxwell}, \citenamefont {Gauguet}, \citenamefont
  {Weatherill}, \citenamefont {Jones},\ and\ \citenamefont
  {Adams}}]{Adams2010}%
  \BibitemOpen
  \bibfield  {author} {\bibinfo {author} {\bibfnamefont {J.~D.}\ \bibnamefont
  {Pritchard}}, \bibinfo {author} {\bibfnamefont {D.}~\bibnamefont {Maxwell}},
  \bibinfo {author} {\bibfnamefont {A.}~\bibnamefont {Gauguet}}, \bibinfo
  {author} {\bibfnamefont {K.~J.}\ \bibnamefont {Weatherill}}, \bibinfo
  {author} {\bibfnamefont {M.~P.~A.}\ \bibnamefont {Jones}}, \ and\ \bibinfo
  {author} {\bibfnamefont {C.~S.}\ \bibnamefont {Adams}},\ }\href {\doibase
  10.1103/PhysRevLett.105.193603} {\bibfield  {journal} {\bibinfo  {journal}
  {Phys. Rev. Lett.}\ }\textbf {\bibinfo {volume} {105}},\ \bibinfo {pages}
  {193603} (\bibinfo {year} {2010})}\BibitemShut {NoStop}%
\bibitem [{\citenamefont {Peyronel}\ \emph {et~al.}(2012)\citenamefont
  {Peyronel}, \citenamefont {Firstenberg}, \citenamefont {Liang}, \citenamefont
  {Hofferberth}, \citenamefont {Gorshkov}, \citenamefont {Pohl}, \citenamefont
  {Lukin},\ and\ \citenamefont {Vuleti\'{c}}}]{Vuletic2012}%
  \BibitemOpen
  \bibfield  {author} {\bibinfo {author} {\bibfnamefont {T.}~\bibnamefont
  {Peyronel}}, \bibinfo {author} {\bibfnamefont {O.}~\bibnamefont
  {Firstenberg}}, \bibinfo {author} {\bibfnamefont {Q.}~\bibnamefont {Liang}},
  \bibinfo {author} {\bibfnamefont {S.}~\bibnamefont {Hofferberth}}, \bibinfo
  {author} {\bibfnamefont {A.~V.}\ \bibnamefont {Gorshkov}}, \bibinfo {author}
  {\bibfnamefont {T.}~\bibnamefont {Pohl}}, \bibinfo {author} {\bibfnamefont
  {M.~D.}\ \bibnamefont {Lukin}}, \ and\ \bibinfo {author} {\bibfnamefont
  {V.}~\bibnamefont {Vuleti\'{c}}},\ }\href {\doibase 10.1038/nature11361}
  {\bibfield  {journal} {\bibinfo  {journal} {Nature}\ }\textbf {\bibinfo
  {volume} {488}},\ \bibinfo {pages} {57} (\bibinfo {year} {2012})}\BibitemShut
  {NoStop}%
\bibitem [{\citenamefont {Fleischhauer}\ \emph {et~al.}(2005)\citenamefont
  {Fleischhauer}, \citenamefont {Imamo\u{g}lu},\ and\ \citenamefont
  {Marangos}}]{Fleischhauer2005}%
  \BibitemOpen
  \bibfield  {author} {\bibinfo {author} {\bibfnamefont {M.}~\bibnamefont
  {Fleischhauer}}, \bibinfo {author} {\bibfnamefont {A.}~\bibnamefont
  {Imamo\u{g}lu}}, \ and\ \bibinfo {author} {\bibfnamefont {J.}~\bibnamefont
  {Marangos}},\ }\href {http://link.aps.org/doi/10.1103/RevModPhys.77.633}
  {\bibfield  {journal} {\bibinfo  {journal} {Rev. Mod. Phys.}\ }\textbf
  {\bibinfo {volume} {77}},\ \bibinfo {pages} {633} (\bibinfo {year}
  {2005})}\BibitemShut {NoStop}%
\bibitem [{\citenamefont {Dudin}\ and\ \citenamefont
  {Kuzmich}(2012)}]{Kuzmich2012b}%
  \BibitemOpen
  \bibfield  {author} {\bibinfo {author} {\bibfnamefont {Y.~O.}\ \bibnamefont
  {Dudin}}\ and\ \bibinfo {author} {\bibfnamefont {A.}~\bibnamefont
  {Kuzmich}},\ }\href {\doibase 10.1126/science.1217901} {\bibfield  {journal}
  {\bibinfo  {journal} {Science}\ }\textbf {\bibinfo {volume} {336}},\ \bibinfo
  {pages} {887} (\bibinfo {year} {2012})}\BibitemShut {NoStop}%
\bibitem [{\citenamefont {Firstenberg}\ \emph {et~al.}(2013)\citenamefont
  {Firstenberg}, \citenamefont {Peyronel}, \citenamefont {Liang}, \citenamefont
  {Gorshkov}, \citenamefont {Lukin},\ and\ \citenamefont
  {Vuleti\'{c}}}]{Vuletic2013b}%
  \BibitemOpen
  \bibfield  {author} {\bibinfo {author} {\bibfnamefont {O.}~\bibnamefont
  {Firstenberg}}, \bibinfo {author} {\bibfnamefont {T.}~\bibnamefont
  {Peyronel}}, \bibinfo {author} {\bibfnamefont {Q.}~\bibnamefont {Liang}},
  \bibinfo {author} {\bibfnamefont {A.~V.}\ \bibnamefont {Gorshkov}}, \bibinfo
  {author} {\bibfnamefont {M.~D.}\ \bibnamefont {Lukin}}, \ and\ \bibinfo
  {author} {\bibfnamefont {V.}~\bibnamefont {Vuleti\'{c}}},\ }\href
  {http://dx.doi.org/10.1038/nature12512} {\bibfield  {journal} {\bibinfo
  {journal} {Nature}\ }\textbf {\bibinfo {volume} {502}},\ \bibinfo {pages}
  {71} (\bibinfo {year} {2013})}\BibitemShut {NoStop}%
\bibitem [{\citenamefont {Li}\ \emph {et~al.}(2013)\citenamefont {Li},
  \citenamefont {Dudin},\ and\ \citenamefont {Kuzmich}}]{Kuzmich2013}%
  \BibitemOpen
  \bibfield  {author} {\bibinfo {author} {\bibfnamefont {L.}~\bibnamefont
  {Li}}, \bibinfo {author} {\bibfnamefont {Y.~O.}\ \bibnamefont {Dudin}}, \
  and\ \bibinfo {author} {\bibfnamefont {A.}~\bibnamefont {Kuzmich}},\ }\href
  {\doibase 10.1038/nature12227} {\bibfield  {journal} {\bibinfo  {journal}
  {Nature}\ }\textbf {\bibinfo {volume} {498}},\ \bibinfo {pages} {466}
  (\bibinfo {year} {2013})}\BibitemShut {NoStop}%
\bibitem [{\citenamefont {Baur}\ \emph {et~al.}(2014)\citenamefont {Baur},
  \citenamefont {Tiarks}, \citenamefont {Rempe},\ and\ \citenamefont
  {D\"urr}}]{Duerr2014}%
  \BibitemOpen
  \bibfield  {author} {\bibinfo {author} {\bibfnamefont {S.}~\bibnamefont
  {Baur}}, \bibinfo {author} {\bibfnamefont {D.}~\bibnamefont {Tiarks}},
  \bibinfo {author} {\bibfnamefont {G.}~\bibnamefont {Rempe}}, \ and\ \bibinfo
  {author} {\bibfnamefont {S.}~\bibnamefont {D\"urr}},\ }\href {\doibase
  10.1103/PhysRevLett.112.073901} {\bibfield  {journal} {\bibinfo  {journal}
  {Phys. Rev. Lett.}\ }\textbf {\bibinfo {volume} {112}},\ \bibinfo {pages}
  {073901} (\bibinfo {year} {2014})}\BibitemShut {NoStop}%
\bibitem [{\citenamefont {Gorshkov}\ \emph {et~al.}(2011)\citenamefont
  {Gorshkov}, \citenamefont {Otterbach}, \citenamefont {Fleischhauer},
  \citenamefont {Pohl},\ and\ \citenamefont {Lukin}}]{Lukin2011}%
  \BibitemOpen
  \bibfield  {author} {\bibinfo {author} {\bibfnamefont {A.~V.}\ \bibnamefont
  {Gorshkov}}, \bibinfo {author} {\bibfnamefont {J.}~\bibnamefont {Otterbach}},
  \bibinfo {author} {\bibfnamefont {M.}~\bibnamefont {Fleischhauer}}, \bibinfo
  {author} {\bibfnamefont {T.}~\bibnamefont {Pohl}}, \ and\ \bibinfo {author}
  {\bibfnamefont {M.~D.}\ \bibnamefont {Lukin}},\ }\href {\doibase
  10.1103/PhysRevLett.107.133602} {\bibfield  {journal} {\bibinfo  {journal}
  {Phys. Rev. Lett.}\ }\textbf {\bibinfo {volume} {107}},\ \bibinfo {pages}
  {133602} (\bibinfo {year} {2011})}\BibitemShut {NoStop}%
\bibitem [{\citenamefont {G\"unter}\ \emph {et~al.}(2012)\citenamefont
  {G\"unter}, \citenamefont {{Robert-de-Saint-Vincent}}, \citenamefont
  {Schempp}, \citenamefont {Hofmann}, \citenamefont {Whitlock},\ and\
  \citenamefont {Weidem\"uller}}]{Weidemueller2012}%
  \BibitemOpen
  \bibfield  {author} {\bibinfo {author} {\bibfnamefont {G.}~\bibnamefont
  {G\"unter}}, \bibinfo {author} {\bibfnamefont {M.}~\bibnamefont
  {{Robert-de-Saint-Vincent}}}, \bibinfo {author} {\bibfnamefont
  {H.}~\bibnamefont {Schempp}}, \bibinfo {author} {\bibfnamefont {C.~S.}\
  \bibnamefont {Hofmann}}, \bibinfo {author} {\bibfnamefont {S.}~\bibnamefont
  {Whitlock}}, \ and\ \bibinfo {author} {\bibfnamefont {M.}~\bibnamefont
  {Weidem\"uller}},\ }\href {\doibase 10.1103/PhysRevLett.108.013002}
  {\bibfield  {journal} {\bibinfo  {journal} {Phys. Rev. Lett.}\ }\textbf
  {\bibinfo {volume} {108}},\ \bibinfo {pages} {013002} (\bibinfo {year}
  {2012})}\BibitemShut {NoStop}%
\bibitem [{\citenamefont {Olmos}\ \emph {et~al.}(2011)\citenamefont {Olmos},
  \citenamefont {Li}, \citenamefont {Hofferberth},\ and\ \citenamefont
  {Lesanovsky}}]{Lesanovsky2011}%
  \BibitemOpen
  \bibfield  {author} {\bibinfo {author} {\bibfnamefont {B.}~\bibnamefont
  {Olmos}}, \bibinfo {author} {\bibfnamefont {W.}~\bibnamefont {Li}}, \bibinfo
  {author} {\bibfnamefont {S.}~\bibnamefont {Hofferberth}}, \ and\ \bibinfo
  {author} {\bibfnamefont {I.}~\bibnamefont {Lesanovsky}},\ }\href {\doibase
  10.1103/PhysRevA.84.041607} {\bibfield  {journal} {\bibinfo  {journal} {Phys.
  Rev. A}\ }\textbf {\bibinfo {volume} {84}},\ \bibinfo {pages} {041607}
  (\bibinfo {year} {2011})}\BibitemShut {NoStop}%
\bibitem [{\citenamefont {Ji}\ \emph {et~al.}(2013)\citenamefont {Ji},
  \citenamefont {Ates}, \citenamefont {Garrahan},\ and\ \citenamefont
  {Lesanovsky}}]{Lesanovsky2013}%
  \BibitemOpen
  \bibfield  {author} {\bibinfo {author} {\bibfnamefont {S.}~\bibnamefont
  {Ji}}, \bibinfo {author} {\bibfnamefont {C.}~\bibnamefont {Ates}}, \bibinfo
  {author} {\bibfnamefont {J.~P.}\ \bibnamefont {Garrahan}}, \ and\ \bibinfo
  {author} {\bibfnamefont {I.}~\bibnamefont {Lesanovsky}},\ }\href
  {http://stacks.iop.org/1742-5468/2013/i=02/a=P02005} {\bibfield  {journal}
  {\bibinfo  {journal} {Journal of Statistical Mechanics: Theory and
  Experiment}\ }\textbf {\bibinfo {volume} {2013}},\ \bibinfo {pages} {P02005}
  (\bibinfo {year} {2013})}\BibitemShut {NoStop}%
\bibitem [{\citenamefont {Balewski}\ \emph {et~al.}(2013)\citenamefont
  {Balewski}, \citenamefont {Krupp}, \citenamefont {Gaj}, \citenamefont
  {Peter}, \citenamefont {B\"uchler}, \citenamefont {L\"ow}, \citenamefont
  {Hofferberth},\ and\ \citenamefont {Pfau}}]{Pfau2013}%
  \BibitemOpen
  \bibfield  {author} {\bibinfo {author} {\bibfnamefont {J.}~\bibnamefont
  {Balewski}}, \bibinfo {author} {\bibfnamefont {A.}~\bibnamefont {Krupp}},
  \bibinfo {author} {\bibfnamefont {A.}~\bibnamefont {Gaj}}, \bibinfo {author}
  {\bibfnamefont {D.}~\bibnamefont {Peter}}, \bibinfo {author} {\bibfnamefont
  {H.}~\bibnamefont {B\"uchler}}, \bibinfo {author} {\bibfnamefont
  {R.}~\bibnamefont {L\"ow}}, \bibinfo {author} {\bibfnamefont
  {S.}~\bibnamefont {Hofferberth}}, \ and\ \bibinfo {author} {\bibfnamefont
  {T.}~\bibnamefont {Pfau}},\ }\href {\doibase doi:10.1038/nature12592}
  {\bibfield  {journal} {\bibinfo  {journal} {Nature}\ }\textbf {\bibinfo
  {volume} {502}},\ \bibinfo {pages} {664} (\bibinfo {year}
  {2013})}\BibitemShut {NoStop}%
\bibitem [{\citenamefont {van Ditzhuijzen}\ \emph {et~al.}(2008)\citenamefont
  {van Ditzhuijzen}, \citenamefont {Koenderink}, \citenamefont {Hern\'andez},
  \citenamefont {Robicheaux}, \citenamefont {Noordam},\ and\ \citenamefont {van
  Linden van~den Heuvell}}]{vanLinden2008}%
  \BibitemOpen
  \bibfield  {author} {\bibinfo {author} {\bibfnamefont {C.~S.~E.}\
  \bibnamefont {van Ditzhuijzen}}, \bibinfo {author} {\bibfnamefont {A.~F.}\
  \bibnamefont {Koenderink}}, \bibinfo {author} {\bibfnamefont {J.~V.}\
  \bibnamefont {Hern\'andez}}, \bibinfo {author} {\bibfnamefont
  {F.}~\bibnamefont {Robicheaux}}, \bibinfo {author} {\bibfnamefont {L.~D.}\
  \bibnamefont {Noordam}}, \ and\ \bibinfo {author} {\bibfnamefont {H.~B.}\
  \bibnamefont {van Linden van~den Heuvell}},\ }\href {\doibase
  10.1103/PhysRevLett.100.243201} {\bibfield  {journal} {\bibinfo  {journal}
  {Phys. Rev. Lett.}\ }\textbf {\bibinfo {volume} {100}},\ \bibinfo {pages}
  {243201} (\bibinfo {year} {2008})}\BibitemShut {NoStop}%
\bibitem [{\citenamefont {G\"unter}\ \emph {et~al.}(2013)\citenamefont
  {G\"unter}, \citenamefont {Schempp}, \citenamefont
  {{Robert-de-Saint-Vincent}}, \citenamefont {Gavryusev}, \citenamefont
  {Helmrich}, \citenamefont {Hofmann}, \citenamefont {Whitlock},\ and\
  \citenamefont {Weidem\"uller}}]{Weidemueller2013b}%
  \BibitemOpen
  \bibfield  {author} {\bibinfo {author} {\bibfnamefont {G.}~\bibnamefont
  {G\"unter}}, \bibinfo {author} {\bibfnamefont {H.}~\bibnamefont {Schempp}},
  \bibinfo {author} {\bibfnamefont {M.}~\bibnamefont
  {{Robert-de-Saint-Vincent}}}, \bibinfo {author} {\bibfnamefont
  {V.}~\bibnamefont {Gavryusev}}, \bibinfo {author} {\bibfnamefont
  {S.}~\bibnamefont {Helmrich}}, \bibinfo {author} {\bibfnamefont {C.~S.}\
  \bibnamefont {Hofmann}}, \bibinfo {author} {\bibfnamefont {S.}~\bibnamefont
  {Whitlock}}, \ and\ \bibinfo {author} {\bibfnamefont {M.}~\bibnamefont
  {Weidem\"uller}},\ }\href {\doibase 10.1126/science.1244843} {\bibfield
  {journal} {\bibinfo  {journal} {Science}\ }\textbf {\bibinfo {volume}
  {342}},\ \bibinfo {pages} {954} (\bibinfo {year} {2013})}\BibitemShut
  {NoStop}%
\bibitem [{\citenamefont {Ebert}\ \emph {et~al.}(2014)\citenamefont {Ebert},
  \citenamefont {Gill}, \citenamefont {Gibbons}, \citenamefont {Zhang},
  \citenamefont {Saffman},\ and\ \citenamefont {Walker}}]{Walker2014}%
  \BibitemOpen
  \bibfield  {author} {\bibinfo {author} {\bibfnamefont {M.}~\bibnamefont
  {Ebert}}, \bibinfo {author} {\bibfnamefont {A.}~\bibnamefont {Gill}},
  \bibinfo {author} {\bibfnamefont {M.}~\bibnamefont {Gibbons}}, \bibinfo
  {author} {\bibfnamefont {X.}~\bibnamefont {Zhang}}, \bibinfo {author}
  {\bibfnamefont {M.}~\bibnamefont {Saffman}}, \ and\ \bibinfo {author}
  {\bibfnamefont {T.~G.}\ \bibnamefont {Walker}},\ }\href {\doibase
  10.1103/PhysRevLett.112.043602} {\bibfield  {journal} {\bibinfo  {journal}
  {Phys. Rev. Lett.}\ }\textbf {\bibinfo {volume} {112}},\ \bibinfo {pages}
  {043602} (\bibinfo {year} {2014})}\BibitemShut {NoStop}%
\bibitem [{\citenamefont {Bochmann}\ \emph {et~al.}(2010)\citenamefont
  {Bochmann}, \citenamefont {M\"ucke}, \citenamefont {Guhl}, \citenamefont
  {Ritter}, \citenamefont {Rempe},\ and\ \citenamefont {Moehring}}]{Rempe2010}%
  \BibitemOpen
  \bibfield  {author} {\bibinfo {author} {\bibfnamefont {J.}~\bibnamefont
  {Bochmann}}, \bibinfo {author} {\bibfnamefont {M.}~\bibnamefont {M\"ucke}},
  \bibinfo {author} {\bibfnamefont {C.}~\bibnamefont {Guhl}}, \bibinfo {author}
  {\bibfnamefont {S.}~\bibnamefont {Ritter}}, \bibinfo {author} {\bibfnamefont
  {G.}~\bibnamefont {Rempe}}, \ and\ \bibinfo {author} {\bibfnamefont {D.~L.}\
  \bibnamefont {Moehring}},\ }\href {\doibase 10.1103/PhysRevLett.104.203601}
  {\bibfield  {journal} {\bibinfo  {journal} {Phys. Rev. Lett.}\ }\textbf
  {\bibinfo {volume} {104}},\ \bibinfo {pages} {203601} (\bibinfo {year}
  {2010})}\BibitemShut {NoStop}%
\bibitem [{\citenamefont {Vlastakis}\ \emph {et~al.}(2013)\citenamefont
  {Vlastakis}, \citenamefont {Kirchmair}, \citenamefont {Leghtas},
  \citenamefont {Nigg}, \citenamefont {Frunzio}, \citenamefont {Girvin},
  \citenamefont {Mirrahimi}, \citenamefont {Devoret},\ and\ \citenamefont
  {Schoelkopf}}]{Schoelkopf2013b}%
  \BibitemOpen
  \bibfield  {author} {\bibinfo {author} {\bibfnamefont {B.}~\bibnamefont
  {Vlastakis}}, \bibinfo {author} {\bibfnamefont {G.}~\bibnamefont
  {Kirchmair}}, \bibinfo {author} {\bibfnamefont {Z.}~\bibnamefont {Leghtas}},
  \bibinfo {author} {\bibfnamefont {S.~E.}\ \bibnamefont {Nigg}}, \bibinfo
  {author} {\bibfnamefont {L.}~\bibnamefont {Frunzio}}, \bibinfo {author}
  {\bibfnamefont {S.~M.}\ \bibnamefont {Girvin}}, \bibinfo {author}
  {\bibfnamefont {M.}~\bibnamefont {Mirrahimi}}, \bibinfo {author}
  {\bibfnamefont {M.~H.}\ \bibnamefont {Devoret}}, \ and\ \bibinfo {author}
  {\bibfnamefont {R.~J.}\ \bibnamefont {Schoelkopf}},\ }\href {\doibase
  10.1126/science.1243289} {\bibfield  {journal} {\bibinfo  {journal}
  {Science}\ }\textbf {\bibinfo {volume} {342}},\ \bibinfo {pages} {607}
  (\bibinfo {year} {2013})}\BibitemShut {NoStop}%
\bibitem [{\citenamefont {Honer}\ \emph {et~al.}(2011)\citenamefont {Honer},
  \citenamefont {L\"ow}, \citenamefont {Weimer}, \citenamefont {Pfau},\ and\
  \citenamefont {B\"uchler}}]{Buechler2011}%
  \BibitemOpen
  \bibfield  {author} {\bibinfo {author} {\bibfnamefont {J.}~\bibnamefont
  {Honer}}, \bibinfo {author} {\bibfnamefont {R.}~\bibnamefont {L\"ow}},
  \bibinfo {author} {\bibfnamefont {H.}~\bibnamefont {Weimer}}, \bibinfo
  {author} {\bibfnamefont {T.}~\bibnamefont {Pfau}}, \ and\ \bibinfo {author}
  {\bibfnamefont {H.~P.}\ \bibnamefont {B\"uchler}},\ }\href {\doibase
  10.1103/PhysRevLett.107.093601} {\bibfield  {journal} {\bibinfo  {journal}
  {Phys. Rev. Lett.}\ }\textbf {\bibinfo {volume} {107}},\ \bibinfo {pages}
  {093601} (\bibinfo {year} {2011})}\BibitemShut {NoStop}%
\bibitem [{\citenamefont {Reiserer}\ \emph {et~al.}(2013)\citenamefont
  {Reiserer}, \citenamefont {Ritter},\ and\ \citenamefont {Rempe}}]{Rempe2013}%
  \BibitemOpen
  \bibfield  {author} {\bibinfo {author} {\bibfnamefont {A.}~\bibnamefont
  {Reiserer}}, \bibinfo {author} {\bibfnamefont {S.}~\bibnamefont {Ritter}}, \
  and\ \bibinfo {author} {\bibfnamefont {G.}~\bibnamefont {Rempe}},\ }\href
  {\doibase 10.1126/science.1246164} {\bibfield  {journal} {\bibinfo  {journal}
  {Science}\ }\textbf {\bibinfo {volume} {342}},\ \bibinfo {pages} {1349}
  (\bibinfo {year} {2013})}\BibitemShut {NoStop}%
\bibitem [{\citenamefont {B\'eguin}\ \emph {et~al.}(2013)\citenamefont
  {B\'eguin}, \citenamefont {Vernier}, \citenamefont {Chicireanu},
  \citenamefont {Lahaye},\ and\ \citenamefont {Browaeys}}]{Browaeys2013}%
  \BibitemOpen
  \bibfield  {author} {\bibinfo {author} {\bibfnamefont {L.}~\bibnamefont
  {B\'eguin}}, \bibinfo {author} {\bibfnamefont {A.}~\bibnamefont {Vernier}},
  \bibinfo {author} {\bibfnamefont {R.}~\bibnamefont {Chicireanu}}, \bibinfo
  {author} {\bibfnamefont {T.}~\bibnamefont {Lahaye}}, \ and\ \bibinfo {author}
  {\bibfnamefont {A.}~\bibnamefont {Browaeys}},\ }\href {\doibase
  10.1103/PhysRevLett.110.263201} {\bibfield  {journal} {\bibinfo  {journal}
  {Phys. Rev. Lett.}\ }\textbf {\bibinfo {volume} {110}},\ \bibinfo {pages}
  {263201} (\bibinfo {year} {2013})}\BibitemShut {NoStop}%
\bibitem [{\citenamefont {{Li}}\ \emph {et~al.}(2014)\citenamefont {{Li}},
  \citenamefont {{Viscor}}, \citenamefont {{Hofferberth}},\ and\ \citenamefont
  {{Lesanovsky}}}]{Lesanovsky2014}%
  \BibitemOpen
  \bibfield  {author} {\bibinfo {author} {\bibfnamefont {W.}~\bibnamefont
  {{Li}}}, \bibinfo {author} {\bibfnamefont {D.}~\bibnamefont {{Viscor}}},
  \bibinfo {author} {\bibfnamefont {S.}~\bibnamefont {{Hofferberth}}}, \ and\
  \bibinfo {author} {\bibfnamefont {I.}~\bibnamefont {{Lesanovsky}}},\
  }\href@noop {} {\bibfield  {journal} {\bibinfo  {journal} {ArXiv e-prints}\ }
  (\bibinfo {year} {2014})},\ \Eprint {http://arxiv.org/abs/1404.0311}
  {arXiv:1404.0311 [physics.atom-ph]} \BibitemShut {NoStop}%
\bibitem [{\citenamefont {He}\ \emph {et~al.}(2014)\citenamefont {He},
  \citenamefont {Sharypov}, \citenamefont {Sheng}, \citenamefont {Simon},\ and\
  \citenamefont {Xiao}}]{Xiao2014}%
  \BibitemOpen
  \bibfield  {author} {\bibinfo {author} {\bibfnamefont {B.}~\bibnamefont
  {He}}, \bibinfo {author} {\bibfnamefont {A.~V.}\ \bibnamefont {Sharypov}},
  \bibinfo {author} {\bibfnamefont {J.}~\bibnamefont {Sheng}}, \bibinfo
  {author} {\bibfnamefont {C.}~\bibnamefont {Simon}}, \ and\ \bibinfo {author}
  {\bibfnamefont {M.}~\bibnamefont {Xiao}},\ }\href {\doibase
  10.1103/PhysRevLett.112.133606} {\bibfield  {journal} {\bibinfo  {journal}
  {Phys. Rev. Lett.}\ }\textbf {\bibinfo {volume} {112}},\ \bibinfo {pages}
  {133606} (\bibinfo {year} {2014})}\BibitemShut {NoStop}%
\bibitem [{\citenamefont {{Tiarks}}\ \emph {et~al.}(2014)\citenamefont
  {{Tiarks}}, \citenamefont {{Baur}}, \citenamefont {{Schneider}},
  \citenamefont {{D{\"u}rr}},\ and\ \citenamefont {{Rempe}}}]{Rempe2014}%
  \BibitemOpen
  \bibfield  {author} {\bibinfo {author} {\bibfnamefont {D.}~\bibnamefont
  {{Tiarks}}}, \bibinfo {author} {\bibfnamefont {S.}~\bibnamefont {{Baur}}},
  \bibinfo {author} {\bibfnamefont {K.}~\bibnamefont {{Schneider}}}, \bibinfo
  {author} {\bibfnamefont {S.}~\bibnamefont {{D{\"u}rr}}}, \ and\ \bibinfo
  {author} {\bibfnamefont {G.}~\bibnamefont {{Rempe}}},\ }\href@noop {}
  {\bibfield  {journal} {\bibinfo  {journal} {ArXiv e-prints}\ } (\bibinfo
  {year} {2014})},\ \Eprint {http://arxiv.org/abs/1404.3061} {arXiv:1404.3061
  [quant-ph]} \BibitemShut {NoStop}%
\end{thebibliography}%
